\documentclass{article}
\usepackage{amsfonts,amssymb, amsmath,mathrsfs}

\newtheorem{prop}{Proposition}

\newtheorem{remark}{Remark}

\def\nn{ \nonumber }
\def\bq{ \begin{equation} }
\def\eq{ \end{equation} }
\def\ben{ \begin{eqnarray} }
\def\en{ \end{eqnarray} }

\def\frac#1#2{\genfrac{}{}{}{}{#1}{#2}}

\def\dfrac#1#2{\displaystyle{\genfrac{}{}{}{}{#1}{#2}}}

\def\ii{{\rm i}}

\begin{document}
%%%%%%%%%%%% TITLE %%%%%%%%%%%%%%

\title{On one  integrable system with a cubic first
integral }

\author{A. V. Vershilov and A. V. Tsiganov \\
\it\small
St.Petersburg State University, St.Petersburg, Russia\\
\it\small e--mail: alexander.vershilov@gmail.com\,, \,\,\,
  andrey.tsiganov@gmail.com}

 \date{}
\maketitle

\vskip0.1truecm
%%%%%%%% A B S T R A C T %%%%%%%%%
\begin{abstract}
Recently  one integrable model with a cubic first integral of motion has been  studied by Valent using
some special coordinate system. We describe the  bi-Hamiltonian structures and
variables of separation for this system.
\end{abstract}

\section{Introduction.}
\setcounter{equation}{0}
The aim of this note is to consider one particular two-dimensional integrable model defined by
natural Hamilton function
\bq \label{nat-h}
H=T+V=\sum_{i,j=1}^2 \mathrm g_{ij}(q_2)\,p_i\,p_j+V(q_1, q_2)\,
\eq
with the metric depending on one variable, and cubic additional integral of motion with the  leading terms
\bq\label{q-def}
H_2= p\, p_1^3+2q\,T\,p_1+\ldots\,,\qquad  p\in\mathbb R\,,\qquad  q\geq 0\,.
\eq
According to \cite{val10},   relevant metrics are described by a finite
number of parameters and lead to a large class of models mainly on the manifolds $\mathbb S^2$ and
$\mathbb H^2$. By suitable choices of the parameters entering the construction these systems are globally defined
and contain as special cases the  known systems of Goryachev-Chaplygin, of Goryachev, and of Dullin and Matveev,  see \cite{val10}.

In \cite{ts11s}  we introduce a natural Poisson bivector depending on two arbitrary functions, which allows us to describe   similar  family of integrable system with  cubic additional integral of motion in spherical coordinates. Below we rewrite this bivector in the coordinate system  used in  \cite{val10},  calculate the corresponding variables of separation and prove that equations of motion are linearized on genus three non-hyperelliptic curve.

\section{ Settings}
\setcounter{equation}{0}
In this section we recall some  necessary facts about  natural bi-integrable  systems on Riemannian manifolds  admitting separation of variables in the Hamilton-Jacobi equation \cite{imm00,ts10,ts11s}

Let $Q$ be a $n$-dimensional  Riemannian manifold.  Its  cotangent bundle $T^*Q $ is naturally
 endowed with canonical invertible Poisson bivector $P$, which  has a standard form in fibered coordinates $z=(q_1,\ldots,q_n,p_1,\ldots,p_n)$ on  $T^*Q$
 \bq \label{p-can}
P=\left(
  \begin{array}{cc}
    0 & \mathrm{I} \\
    -\mathrm{I} & 0
      \end{array}
\right)\,,\qquad \{f,g\}=\langle P\,df,dg\rangle=\sum_{i=1}^{2n}P_{ij}\dfrac{\partial f}{\partial z_i}\dfrac{\partial g}{\partial z_j}\,,
\eq
In order to calculate the variables of separation  for the given integrable system with integrals of motion $H_1,\ldots,H_n$ in involution
\[
\{H_i,H_j\}=0\,,\qquad i,j=1,\ldots,n,
\]
in the bi-Hamiltonian set-up we have to solve equations
\bq\label{comp-p}
[P,P]=[P',P']=0\,,\qquad
\eq
where $[.,.]$ means a Schouten bracket, and
\bq\label{bi-inv}
\{H_i,H_j\}'=0\,,\qquad i,j=1,\ldots,n,\qquad \{f,g\}'=\langle P'\,df,dg\rangle\,,
\eq
with respect to the Poisson bivector $P'$.  Then we have to calculate the so-called Nijenhuis operator (or hereditary, or recursion operator)
\bq\label{rec-op}
N=P' P^{-1}\,.
\eq
If $N$ has, at every point, the maximal number of different functionally independent eigenvalues $u_1,\ldots,u_n$,  they  may be treated either as action variables (integrals of motion) or as variables of separation for this dynamical system \cite{imm00,ts10}.

Separation of variables for natural integrable systems with higher order integrals of motion always involves generic canonical transformation of the whole phase space. The definition of the natural Hamiltonians (\ref{nat-h}) is non-invariant with respect to such  transformations of coordinates on the whole phase space (cotangent bundle).

In the situation, when habitual objects (geodesics, metrics and potentials) lose their geometric sense, and the remaining invariant equations (\ref{comp-p},\ref{bi-inv}) have infinite number of solutions, the notion of the natural Poisson bivectors becames de-facto a very useful practical tool for the calculation of variables of separation  \cite{ts10,ts11s}. It is an experimental fact supported by all the known constructions of the variables of separation on the sphere. We try to draw attention to this experimental fact in order to find a suitable geometric explanation of this phenomenon.

Similar to the natural Hamilton function on $T^*Q $, natural  Poisson bivector $P'$  is a sum of
the geodesic Poisson bivector $P'_T$ and the  potential Poisson bivector
defined by a torsionless (1,1) tensor field $\Lambda(q_1,\ldots, q_n)$ on $Q$ associated with potential $V$ \cite{ts10,ts11s}:
\bq\label{n-p}
P'= P'_T+\left(
      \begin{array}{cc}
        0 & \Lambda_{ij} \\
        \\
         -\Lambda_{ji}\quad &\displaystyle \sum_{k=1}^n\left(\dfrac{\partial \Lambda_{ki}}{\partial q_j}-\dfrac{\partial \Lambda_{kj}}{\partial q_i}\right)p_k
      \end{array}
    \right)\,.
\eq
The  geodesic Poisson bivector $P'_T$ is  defined by $n\times n$ geodesic matrix $\Pi$ on $T^*Q$ :
\bq\label{p2-sph2}
 P'_T=
 \left(
      \begin{array}{cc}
        \displaystyle \sum_{k=1}^n \mathrm{x}_{jk}(q)\dfrac{\partial \Pi_{jk}}{\partial p_i}-\mathrm{y}_{ik}(q)\dfrac{\partial \Pi_{ik}}{\partial p_j} & \Pi_{ij} \\
        \\
         -\Pi_{ji}\quad&\displaystyle \sum_{k=1}^n\left(\dfrac{\partial \Pi_{ki}}{\partial q_j}-\dfrac{\partial \Pi_{kj}}{\partial q_i}\right)\,\mathrm z_{k}(p\,)\\
      \end{array}
    \right)
\,.
 \eq
In fact, for  the given matrix $\Pi$ functions $\mathrm{x,y}$ and $\mathrm z$ are completely determined by the equations
\bq\label{comp-pt}
[P,P'_T]=[P'_T,P'_T]=0\,,
\eq
whereas $\Lambda$ is obtained as a solution of the equation (\ref{comp-p}). Discussion of this  useful anzats (\ref{n-p}) may be found in \cite{ts10,ts11s}.

\subsection{Two-dimensional integrable systems with cubic integrals of motion}
Let $Q$ be a $2$-dimensional  Riemannian manifold, and its  cotangent bundle $T^*Q $ is naturally
 endowed with four  canonical coordinates $q_{1,2}$ and $p_{1,2}$.

One of the  natural Poisson  bivectors $P'$ (\ref{n-p})  listed in  \cite{ts11s}  is defined by  geodesic matrix
depending on one variable $q_2$:
 \bq\label{gor-pi}
 \Pi=\left(
 \begin{array}{cc}
 0 & -\dfrac{\mathrm i}{2}\,\left(\dfrac{\partial }{\partial q_2}+\dfrac{2h(q_2)}{g(q_2)} \right)\,\mathbb F\\ \\
 0 & \mathbb F
 \end{array}
 \right)\,,\qquad \mathbb F=\Bigl(g(q_2)p_2-\mathrm i h(q_2)p_1\Bigr)^2\,,\quad \mathrm i=\sqrt{-1}\,,
 \eq
and diagonal potential matrix
\bq\label{gor-l}
\Lambda=\alpha\,\exp\left(\mathrm i q_1-\int\dfrac{h(q_2)}{g(q_2)}\,d q_2\right)\left(
 \begin{array}{cc}
 1 & 0 \\ \\
 0 & 1
 \end{array}
 \right)\,,\qquad \alpha\in\mathbb R\,.
\eq
Here  $g(q_2)$ and $h(q_2)$ are arbitrary functions and
\[
\mathrm x_{22}=-\dfrac{g(q_2)}{2h(q_2)}\,.
\]
Other functions $\mathrm y_{ik}$, $\mathrm x_{ik}$ and $\mathrm z_k$ in the definition (\ref{p2-sph2})  equal  zero.

A characteristic polynomial of the corresponding recursion operator $N=P'P^{-1}$ reads as
\[
\mathbb B(\eta)=\det (N-\eta\mathrm I)=\Bigl(\eta^2-(\mathbb F+2\Lambda_{11})\,\eta+\Lambda_{11}^2\Bigr)^2\,.
\]
Here  $\mathbb F$  (\ref{gor-pi}) is a complete square, it allows us to introduce  the coordinates  $v_{1,2}$ as
 roots of the linear in momenta $p_{1,2}$ polynomial
\bq\label{q-gor}
 B(\lambda)=(\lambda-v_1)(\lambda-v_2)=
\lambda^2-\ii \sqrt{ \mathbb F }\lambda+\Lambda_{11}\,.
\eq
In this coordinates $\mathbb B=(\eta+v_1^2)^2(\eta+v_2^2)^2$.  We prefer to use the linear in momenta polynomial $B(\lambda)$ instead of $\mathbb B(\eta)$ because in this case  it is easy to find a solution of the equations
\bq\label{ab-eq}
\{ B(\lambda), A(\mu)\}=\dfrac{\lambda}{\mu-\lambda}
\left(\dfrac{ B(\lambda)}{\lambda}-\dfrac{ B(\mu)}{\mu}\right)\,,
\qquad \{ A(\lambda), A(\mu)\}=0\,,\eq
with respect to other  linear in momenta polynomial
\[
 A(\lambda)=\int\dfrac{\mathrm idq_2}{{g(q_2)}}-\dfrac{\mathrm ip_1}{\lambda}\,.
\]
Equations (\ref{ab-eq}) entail that canonically conjugated   momenta  are equal to
\bq\label{p-gor}
p_{v_{1,2}}= A(\lambda=v_{1,2})\,.
\eq
Substituting  variables
\[
x=a_0\,v^{-1}_{k}\,,\qquad z=\ii p_{k}\,,\qquad k=1,2,\qquad a_0\in\mathbb R\,,
\]
into the generic equation of the so-called  (3,4) algebraic curve \cite{en07}
\bq\label{g-Phi}
\Phi(z,x)=z^3+(a_1x+a_2)z^2+(H_1x^2+b_1x+b_2)z+x^4+H_2x^3+c_1x^2+c_2x+c_3=0\,,\qquad a_k,b_k.c_k\in\mathbb R\,,
\eq
and solving  the resulting equations with respect to $H_{1,2}$,
one gets the following  Hamilton function
\bq\label{g-Ham}
H_1=T+V
+\dfrac{1}{a_0w_2}\Bigl(
(a_1w_2^2-b_1w_2+c_2)h-2a_1w_2+b_1
\Bigr)p_1+\dfrac{\ii g(a_1w_2^2-b_1w_2+c_2)}{a_0w_2}p_2\,,
\eq
where geodesic Hamiltonian and potential are equal to
\ben
T&=&\dfrac{1}{a_0^2w_2}\Bigl(
(c_3-w_2^3+a_2w_2^2-b_2w_2)h^2+(b_2+3w_2^2-2a_2w_2)h-3w_2+a_2
\Bigr)p_1^2\nn\\
\nn\\
&+&\dfrac{\ii g}{a_0^2w_2}\Bigl(
2(c_3-w_2^3+a_2w_2^2-b_2w_2)h+b_2+3w_2^2-2a_2w_2
\Bigr)\,p_1p_2
-\dfrac{g^2(c_3-w_2^3+a_2w_2^2-b_2w_2)}{a_0^2w_2}\,p_2^2\nn\\
\nn\\
V&=&-\dfrac{a_0^2\mathrm e^{-\ii q_1}}{\alpha w_1w_2}-\dfrac{\alpha w_1(c_3-w_2^3+a_2w_2^2-b_2w_2)\mathrm e^{\ii q_1}}{a_0^2w_2}+\dfrac{c_1}{w_2}\,.\nn
\en
Here
\[
w_1=\exp\left(-\int\dfrac{h(q_2)}{g(q_2)}\,dq_2\right) \,,\qquad w_2=\int\dfrac{dq_2}{g(q_2)}\,.
\]
Second integral of motion $H_2$ is a cubic polynomial in momenta $p_{1,2}$.

If $h\neq 0$ and $g\neq0$, in order to reduce  Hamiltonian $H_1$ (\ref{g-Ham}) to the natural form, we have to solve two integral equations
\bq\label{hg-eq}
a_1w_2^2-b_1w_2+c_2=0\,,\qquad b_1 -2a_1w_2=0\,,
\eq
with respect to functions $h(q_2)$, $g(q_2)$ and parameters $a_1,b_1,c_2$.
If we want to obtain diagonal metric, then we have to add one more equation to this system.
There is an additional freedom  related with a possible canonical transformation $p_2\to p_2+f(q_2)$, which change linear in momenta terms in (\ref{g-Ham}).

If $Q=\mathbb S^2$  is a two-dimensional unit sphere with  spherical coordinates
 \bq
 q=(q_{1},q_{2})=(\phi,\theta)\qquad\mbox{and}\qquad p=(p_1,p_2)=(p_\phi,p_\theta)\,,\label{p12-coord}
 \eq
then  we could get a whole  family of natural integrable systems on the sphere using  different functions $h(\theta)$, $g(\theta)$ labeled by particular values of parameters in (\ref{hg-eq}) \cite{ts11s,ts09v}.

\section{Natural Poisson bivectors  in $\zeta$-variables }
 In \cite{val10}, solving the equation
\[
\{H_1,H_2\}=0\,
\]
 in framework of the Laplace method,  author prefers to use special canonical coordinates
\bq\label{z-var}
q_1=\phi,\qquad q_2=\zeta,\qquad p_1=p_\phi,\qquad p_2=p_\zeta\,,
\eq
Here $H_1$ is a natural Hamilton function (\ref{nat-h}) with diagonal metric $\mathrm g(\zeta)$ and $H_2$ is fixed by (\ref{q-def}).

 According to \cite{val10},  special case $q = 0$ in (\ref{q-def}) is rather difficult to obtain as the limit of the general case $q\neq 0$. So,  we first integrate the special case at $q = 0$ and only then we consider the generic case.
 In contrast with \cite{val10}, we restrict ourselves only  by local analysis.
\subsection{Case $q=0$}
Let us take integrals of motion from  \cite{val10}, Theorem 1 formulae (14-15):
\ben
H_1^{(0)}&=& \dfrac{1}{2}\left(F\,p_\zeta^2+\dfrac{G}{4F}\,p_\phi^2\right)+\lambda\sqrt{F}\,\cos\phi+\mu\zeta\,,
\qquad ('=D_\zeta)
\nn\\
\label{int-q0}\\
H_2^{(0)}&=&p_\phi^3-2\lambda\left(
\sqrt{F}\sin\phi p_\zeta+(\sqrt{F})'\cos\phi p_\phi\right)-2\mu p_\phi\,,\nn
\en
with
\bq\label{F-val}
F=-2\rho_0+3c_0\zeta+\zeta^3\,,\qquad\qquad G=9c_0^2+24\rho_0\zeta-18c_0\zeta^2-3\zeta^4\,.
\eq
Here $\lambda,\mu,\rho_0$ and $c_0$ are arbitrary parameters.

 Substituting new variables (\ref{z-var}) into  definitions (\ref{gor-pi},\ref{gor-l}) one gets a bivector $P'$ depending on two arbitrary functions $h(\zeta)$ and $g(\zeta)$. Using this bivector we can easily solve the equation (\ref{bi-inv}) and prove the following proposition.

 \begin{prop}
Integrals of motion $H_{1,2}^{(0)}$ (\ref{int-q0}) are in bi-involution
\bq\label{bi-inv2}
\{H_1^{(0)},H_2^{(0)}\}=\{H_1^{(0)},H_2^{(0)}\}'=0
\eq
with respect to a pair of  compatible Poisson brackets associated with the canonical Poisson bivector $P$ (\ref{p-can})
and natural Poisson bivector $P'$ (\ref{n-p}) defined by
\bq \label{pi-val0}
{\Pi}=\left(
      \begin{array}{cc}
        0 & \dfrac{\ii}{2}\left(\dfrac{\partial}{\partial\zeta}+\dfrac{F'}{F}\right){\mathbb F} \\
        \\
        0 & \widetilde{\mathbb F}
      \end{array}
    \right)\,,\qquad{\mathbb F}=-\left(2p_\zeta-\dfrac{\ii F'}{F}\,p_\phi\right)^2
\eq
and
\bq \label{l-val0}
{L}=-\dfrac{4\lambda\mathrm e^{-\ii\phi}}{\sqrt{F}} \left(
          \begin{array}{cc}
         1 & 0 \\
            0 & 1
          \end{array}
        \right)\,,\qquad
{x}_{22} = -\dfrac{F}{F'}\,.
\eq
\end{prop}
As a consequence,  substituting $g(\zeta)=2$ and $h(\zeta)=(\ln F)'$ into variables  $v_{1,2}$ (\ref{q-gor}) and $p_{v_{1,2}}$ (\ref{p-gor}) one gets variables of separation for this integrable model.

Below, in order to extend the known palette of natural Poisson bivectors  listed in \cite{ts11s}, we  consider another solution of the equation (\ref{comp-p}) depending on  both variables $\phi$ and $\zeta$.
 \begin{prop}
Integrals of motion $H_{1,2}^{(0)}$ (\ref{int-q0}) are in the bi-involution (\ref{bi-inv2})
with respect to  canonical Poisson bracket  and  bracket $\{.,\}'$ associated with the natural Poisson bivector $\widetilde{P}'$ (\ref{n-p}) defined by  $2\times 2$ geodesic matrix
\bq\label{pi-val}
\widetilde{\Pi}=\left(
  \begin{array}{cc}
\widetilde{\mathbb F}&-\dfrac{\ii\widetilde{ \mathbb F}'}{2}\\
\\
 0  & 0 \\
  \end{array}
\right)\,,\qquad\qquad \widetilde{\mathbb F}=
\left[\mathrm e^{\ii \phi}\left(\dfrac{\ii F'}{2\sqrt{F}}\,p_\phi+\sqrt{F}\,p_\zeta\right)\right]^2\,,
\eq
diagonal potential matrix
\bq \label{l-val}
\widetilde{\Lambda}=  \lambda \mathrm e^{\ii\phi}\sqrt{F}\left(
          \begin{array}{cc}
         1 & 0 \\
            0 & 1
          \end{array}
        \right)
\eq
and  function
\[
\mathrm y_{11} = -\dfrac{\ii}{2}\,.
\]
Other functions $\mathrm y_{ik}$, $\mathrm x_{ik}$ and $\mathrm z_k$ in the definition (\ref{p2-sph2})  equal  zero.
\end{prop}
As for the Lagrange top, this geodesic matrix (\ref{pi-val}) is factorized $\widetilde{\Pi}=\mathrm e^{2\ii\phi} \, \widehat{\Pi}(\zeta)$, see  discussion in  \cite{ts11s}.

 We are now  able to analyze the corresponding recursion operator $N=\widetilde{P}'P^{-1}$. For instance, we can define the variables of separation $u_{1,2}$ as roots of the  linear in momenta polynomial
\ben\label{u-var}
\widetilde{B}(\eta)&=&(\eta-u_1)(\eta-u_2)=\eta^2-\ii\sqrt{\mathbb F}\,\eta+\Lambda_{11} \\
\nn\\
&=&\eta^2-\ii\mathrm e^{\ii\phi}
\left(
\dfrac{3\ii(c_0+\zeta^2)}{2\sqrt{-2\rho_0+3c_0\zeta+\zeta^3}}\,p_\phi+\sqrt{-2\rho_0+3c_0\zeta+\zeta^3}\,p_\zeta\right)
\eta+\lambda\sqrt{-2\rho_0+3c_0\zeta+\zeta^3}\mathrm e^{\ii\phi}\,,\nn
\en
so that the characteristic polynomial of $N$ reads as
$\det(N-\eta\mathrm I)=(\eta-u_1^2)^2(\eta-u_2^2)^2$. The conjugated momenta are equal to
\[
p_{u_k}=\dfrac{\ii\,p_\phi}{u_k}-\dfrac{\ii \lambda\zeta}{u_k^2}\,,\qquad k=1,2.
\]
 The inverse transformation looks like
\bq\label{inv-tr0}
\begin{array}{ll}
\zeta=-\dfrac{\ii u_1u_2(u_1p_{u_1}-u_2p_{u_2})}{\lambda(u_1-u_2)}\,,\qquad&
p_\phi=\dfrac{\ii(u_1^2p_{u_1}-u_2^2p_{u_2})}{u_1-u_2}\,,\\
\\
\phi = \dfrac{\ii}{2}\ln\left(\dfrac{\lambda^2\,F}{u_1^2u_2^2}\right)\,,\qquad& p_\zeta
=-\dfrac{\ii F'}{2F}\,p_\phi-\dfrac{\ii \lambda(u_1+u_2)}{u_1u_2}\,.
\end{array}
\eq
 In these variables of separation our initial  bivector $P'$ (\ref{pi-val0},\ref{l-val0}) looks like
\[
{P}'={3}\left(
               \begin{array}{cccc}
                 0 & 0 & u_1^2p_{u_1}^2 & 0 \\
                 0 & 0 & 0 & u_2^2p_{u_2}^2 \\
                 -u_1^2p_{u_1}^2  & 0 & 0 & 0 \\
                 0 & -u_2^2p_{u_2}^2 & 0 & 0 \\
               \end{array}
             \right)\,.
\]
Now we can calculate the integrals of motion
\ben
H_1^{(0)}&=& -\dfrac{\ii \mu u_1 u_2 (u_1 p_{u_1}-u_2 p_{u_2})}{\lambda (u_1-u_2)}+\dfrac{(u_2^2+u_1 u_2+u_1^2) \rho_0 \lambda^2}{u_1^2 u_2^2}-\dfrac{3\ii c_0 (u_1 p_{u_2}-p_{u_1} u_2) \lambda}{2(u_1-u_2)}+\dfrac{ u_1 u_2}2\nn\\
\nn\\
&-&\dfrac{\ii (u_1^2 p_{u_1}^2+u_2 p_{u_2} u_1 p_{u_1}+u_2^2 p_{u_2}^2) (u_1 p_{u_1}-u_2 p_{u_2}) u_1 u_2}{2(u_1-u_2) \lambda}\,,\nn\\
\nn\\
H_2^{(0)}&=& -\dfrac{2\ii \mu (p_{u_1} u_1^2-p_{u_2} u_2^2)}{u_1-u_2}+\dfrac{2 (u_1+u_2) \rho_0 \lambda^3}{u_1^2 u_2^2}+\dfrac{3\ii c_0 (p_{u_1}-p_{u_2}) \lambda^2}{u_1-u_2}+\dfrac{u_1+u_2}{\lambda}-\dfrac{\ii (p_{u_1}^3 u_1^4-u_2^4 p_{u_2}^3)}{u_1-u_2}\,.
\nn
\en
and the control matrix
\[
\mathbf F=\left(
            \begin{array}{cc}
              0 & \dfrac{u_1u_2}{2\lambda} \\ \\
              -2\lambda & u_1+u_2
            \end{array}
          \right)\,,\qquad\mbox{from the  definition}\qquad P'dH_i^{(0)}=P\sum_{j=1}^2\mathbf F_{ij}\,dH_j^{(0)},\qquad i=1,\ldots,2.
\]
Suitable normalized left eigenvectors of  control matrix $\mathbf F$ form the St\"ackel matrix $\mathbf S$, so
the notion of $\mathbf F$ allows us  to  compute the corresponding separated relations
\[
\sum_{j=1}^2 \mathbf S_{ij}H_j+U_i=0\,,\qquad i=1,2.
\]
Here functions $\mathbf S_{ij}$ and $U_i$ depend only on one pair $(u_i,p_{u_i})$ of  canonical variables of separation \cite{ts99}.

In our case the integrals of motion  and the variables of separation are related via the following separated relations
\bq\label{seprel0}
\Phi(u,z)=z^3+(3\lambda^2c_0-2\mu u^2)z+\lambda u^4-H_2^{(0)}u^3+2\lambda H_1^{(0)}u^2-2\lambda^3\rho_0=0,
\eq
at $u=u_{1,2}$ and $z=\ii u^2_{1,2}p_{1,2}$. Equation  $\Phi(u,z)=0$ defines genus three  non-hyperelliptic  curve with the following base of the holomorphic differentials
\[
\Omega_1= \dfrac{du}{2\mu u^2-3\lambda^2c_0-3z^2}\,,\qquad\Omega_2= \dfrac{udu}{2\mu u^2-3\lambda^2c_0-3z^2}\,,\qquad \Omega_3= \dfrac{zdu}{2\mu u^2-3\lambda^2c_0-3z^2}\,.
\]
According to  \cite{en07} it is the so-called $(3,4)$ trigonal curve. It is easy to prove that the equations of motion  in variables of separation   have the following  form
\[
\dfrac{\dot{u_1}}{2\mu u_1^2-3\lambda^2c_0-3z_1^2}
+\dfrac{\dot{u_2}}{2\mu u_2^2-3\lambda^2c_0-3z_2^2}=\dfrac{\ii}{2\lambda}\,,\qquad z_k=\ii u^2_kp_k\,,
\]
\[
\dfrac{u_1\dot{u_1}}{2\mu u_1^2-3\lambda^2c_0-3z_1^2}
+\dfrac{u_2\dot{u_2}}{2\mu u_2^2-3\lambda^2c_0-3z_2^2}=0\,.
\]
The aforementioned quadratures  in the integral form
\bq\label{ajm-1}
\int^{u_1}_{u_0} \Omega_1+\int^{u_2}_{u_0} \Omega_1=\dfrac{\ii}{2\lambda}\,t+\beta_1\,,\qquad
\int^{u_1}_{u_0} \Omega_2+\int^{u_2}_{u_0} \Omega_2=\beta_2\,,
\eq
represent the Abel-Jacobi map associated to the genus three non-hyperelliptic curve defined by the equation  $\Phi(q,p)=0$ (\ref{seprel0}). In particular it means that instead of $z$ in $\Omega_{1,2}$ (\ref{ajm-1}) we have to substitute  the function on $u$ obtained from  the separated relation (\ref{seprel0}).

\subsection{Case $q\neq0$}
Let us take integrals of motion from  \cite{val10}, Theorem 5 formulae (39-40):
\ben
H_1&=&\dfrac{1}{2\zeta}\left(Fp_\zeta^2+\dfrac{G}{4F}\,p_\phi^2\right)
+\dfrac{\sqrt{F}}{2q\zeta}\,\cos\phi+\dfrac{\beta_0}{2q\zeta}\,,\nn\\
\label{int-q}\\
H_2&=&p\,p_\phi^3+2q\,H_1p_\phi-\sqrt{F}\sin\phi\, p_\zeta
-(\sqrt{F})'\,\cos\phi\, p_\phi \,,\nn
\en
with
\bq\label{F-valg}
F=c_0+ c_1\zeta+c_2\zeta^2+c_3\zeta^3\,,\qquad G={F'}^2-2F F''\,,\qquad
c_3=\dfrac{p}{q}\,,
\eq
where  $\beta_0,c_0,c_1,c_2$ and $p,q$ are arbitrary parameters.

\begin{prop}
After substitution of the  new function $F$ (\ref{F-valg}) into the previous  definitions (\ref{pi-val0},\ref{l-val0}) and (\ref{pi-val},\ref{l-val}) one gets  two compatible  Poisson bivectors  $P'$  and $\widetilde{P}'$, so that
the integrals of motion $H_{1,2}$ (\ref{int-q}) are in the involution with respect to the corresponding  Poisson brackets.
\end{prop}
Other calculations are standard. Namely,  the variables of separation are given by
\ben
\widetilde{B}&=&(\eta-u_1)(\eta-u_2)=\eta^2-\mathrm e^{\ii\phi}\left(\dfrac{\ii F'}{2\sqrt{F}}\,p_\phi+\sqrt{F}\,p_\zeta\right)\eta
-\dfrac{\sqrt{F}\mathrm e^{\ii \phi}}{2q}\,,\label{u-var-q}\\
 p_{u_k}&=&-\dfrac{\ii p_\phi}{u_k}-\dfrac{\zeta}{2qu_k^2}\,,\nn
 \en
 whereas the inverse transformation reads as
\bq\label{inv-tr}
\begin{array}{ll}
\zeta=\dfrac{2q u_1u_2(u_1p_{u_1}-u_2p_{u_2})}{u_1-u_2}\,,\qquad&
p_\phi=\dfrac{\ii(u_1^2p_{u_1}-u_2^2p_{u_2})}{u_1-u_2}\,,\\
\\
\phi = -\dfrac{ \ii}{2}\ln\left(\dfrac{4q^2u_1^2u_2^2}{F}\right)\,,\qquad& p_\zeta=
 -\dfrac{\ii F'}{2F}p_\phi-\dfrac{u_1+u_2}{2qu_1u_2}\,.
\end{array}
\eq
Then we have to  calculate the control matrix $\mathbf F$ and the corresponding  St\"ackel  matrix $\mathbf S$ in order to get the desired separated relations, which   are obtained by substituting  $u=u_{1,2}$ and $z=p\,u_{1,2}^2p_{u_{1,2}}$ in the following equation
\bq\label{seprel-q}
\Phi=
2z^3-c_2\,z^2-\dfrac{(8q^2H_1u^2-c_1)p}{2q}\,z
-p^2u^4-2\ii p^2 H_2u^3-\dfrac{\beta_0p^2u^2}{q}-\dfrac{p^2c_0}{4q^2}=0\,.
\eq
This equation  $\Phi(u,z)=0$ defines the genus three  non-hyperelliptic  curve ( so-called  (3,4)-curve, see  \cite{en07})
 with  the holomorphic differentials
\bq\label{omega-q}
\Omega_1= \dfrac{du}{-2c_3\partial \Phi/\partial z}\,,\qquad\Omega_2= \dfrac{udu}{-2c_3\partial \Phi/\partial z}\,,\qquad \Omega_3= \dfrac{zdu}{-2c_3\partial \Phi/\partial z}\,.
\eq
The desired  quadratures read as
\[
\dfrac{u\dot{u_1}}{8p^2H_1u_1^2-12c_3z_1^2+4c_2c_3z_1-c_1c_3^2}
+\dfrac{u\dot{u_2}}{8p^2H_1u_2^2-12c_3z_2^2+4c_2c_3z_2-c_1c_3^2}=0\,,\qquad z_k=p\, u^2_k\,p_k\,,
\]
and
\[
\dfrac{z_1\dot{u_1}}{8p^2H_1u_1^2-12c_3z_1^2+4c_2c_3z_1-c_1c_3^2}
+\dfrac{z_2\dot{u_2}}{8p^2H_1u_2^2-12c_3z_2^2+4c_2c_3z_2-c_1c_3^2}=-\dfrac{p}{8}\,.
\]
The above quadratures  in the integral form
\bq\label{ajm-2}
\int^{u_1}_{u_0} \Omega_2+\int^{u_2}_{u_0} \Omega_2=\beta_1\,,\qquad
\int^{u_1}_{u_0} \Omega_3+\int^{u_2}_{u_0} \Omega_3=-\dfrac{p}{8}\,t+\beta_2\,,
\eq
represent the Abel-Jacobi map associated to the genus three non-hyperelliptic curve defined by (\ref{seprel-q}).
If we change $u\to\ii u$, that corresponds to transformation $P'\to-P'$, one gets the  equation $\Phi(u,z)$  (\ref{seprel-q}) with  real coefficients, but the coefficient before time variable in (\ref{ajm-1}) becomes imaginary number.

In order to give an explicit theta-functions solution, one could apply the standard  Weierstrass machinery   describing the inversion of the hyperelliptic quadratures  \cite{w94}.  However,  for the genus three trigonal curve such  solution of the Jacobi inversion problem is only  created. Some of the  results referring to Weierstrass theory for the general trigonal curve of genus three may be found in \cite{en07}.

On the other hand,   for  a non-hyperelliptic  Riemann surface of genus three
the moduli spaces of rank two bundles compactifies into a singular  projective variety, which
is closely related to the Kummer variety and which is a quartic hypersurface of $\mathbb P^7$.
We can try to apply  the generic theory of vector bundles on algebraic varieties
to investigation of  this integrable  system and vise versa \cite{van05}.

\subsection{Change of the time}
If the dimension of the configurational space $n$ is less then genus  $g$ of the separated  curve,  then  the Abel map is either lack of uniqueness or degenerate. In this case we are free to choose  $n$ integrals of motion from the
coefficients of the separated curve,  that equivalent to choose the corresponding  time variables.  The Kepler change of the time,  the Liouville transformations, the Maupertuis--Jacobi transformations and the coupling constant metamorphosis are examples of such duality. One of the well-known examples is a duality between harmonic oscillator and the  Kepler system, which was expanded on the quantum case by Schrodinger.

All the known  such transformations are associated only  with the hyperelliptic curves, see  \cite{ts99,ts00}.
Here we want to show one example related with non-hyperelliptic curve. Let us consider the following  equation
\bq\label{seprel-q-t}
\Phi=
2z^3-c_2\,z^2-\dfrac{(8q^2\alpha u^2-c_1)p}{2q}\,z
-p^2u^4-2\ii p^2 \widehat{H}_2u^3+2p^2\widehat{H}_1u^2-\dfrac{p^2c_0}{4q^2}=0\,,
\eq
 which  is related with the initial equation (\ref{seprel-q}) by permutation of coefficients  $\alpha$ and $\widehat{H}_1$
only. If we substitute variables (\ref{u-var-q})   into (\ref{seprel-q-t}) and  solve the resulting pair of separated equations, then one gets the Hamilton function
\ben
\widehat{H}_1&=& \zeta\Bigl( H_1+\alpha \Bigr)+\dfrac{\beta_0}{2q}=\dfrac{1}{2}\left(Fp_\zeta^2+\dfrac{G}{4F}\,p_\phi^2\right)
+\dfrac{c_3\sqrt{F}}{2p}\,\cos\phi-\alpha\zeta\,,\nn\\
\nn\\
F&=&c_0+ c_1\zeta+c_2\zeta^2+c_3\zeta^3\,,\qquad G={F'}^2-2F F'\,.
\en
 Second integral of motion in this case is equal to
\[
\widehat{H}_2=p\,p_\phi^3-\dfrac{F'}{2\sqrt{F}}\,p_\phi\cos\phi-\sqrt{F}p_\zeta\sin\phi+\dfrac{2\alpha\,p}{c_3}\,p_\phi\,.
\]
Integrals $\widehat{H}_{1,2}$ may be obtained from $H_{1,2}^{(0)}$ (\ref{int-q0})  at special values of parameters using canonical transformation  $\zeta\to \zeta-c_2/3c_3$.

Dual St\"ackel systems with Hamiltonians $H_1$ and $\widehat{H}_1$ are related  to each other by a canonical transformation of the extended phase space \cite{ts99,ts00}.  For the both systems trigonal curve defined by (\ref{seprel-q}) or (\ref{seprel-q-t}) has the same holomorphic differentials $\Omega_{k}$ (\ref{omega-q}), but instead of quadratures (\ref{ajm-2}) in the second case we have
\bq\label{ajm-3}
\int^{u_1}_{u_0} \Omega_1+\int^{u_2}_{u_0} \Omega_1=\dfrac{1}{4pc_3}\,t+\beta_1\,,\qquad
\int^{u_1}_{u_0} \Omega_2+\int^{u_2}_{u_0} \Omega_2=\beta_2\,,
\eq
Similar to  hyperelliptic case, this duality (change of the time) is related with the different choice of the holomorphic differentials in the inversion Jacobi problem  \cite{ts99,ts00}.

Using the same permutation of  coefficients  in the generic equation (\ref{g-Phi}) of (3,4)-curve we can obtain the dual Hamiltonian  to more complicated Hamilton function  (\ref{g-Ham}). Other possible generalization  is related with an another pair of differentials in the Abel map:
\begin{enumerate}
  \item case $q=0$ is associated with   $ \Omega_1$ and $\Omega_2$ (\ref{ajm-3});
  \item case $q\neq0$ is associated with   $ \Omega_2$ and $\Omega_3$ (\ref{ajm-2});
  \item one more integrable case is associated with   $ \Omega_1$ and $\Omega_3$.
\end{enumerate}
Some difficulty here is related with the coordinates choice, because we can not directly recognize interesting
physical   models in $\zeta$-variables. So, different possible generalizations of the
Goryachev, Chaplygin, Dullin-Matveev systems and other systems  from \cite{ts11s,ts09v}
associated with the generic (3,4)-curve
 in term of  physical variables on the two-dimensional sphere will be discussed in
a forthcoming publication.

According to \cite{ts11s},  there are natural generalizations of matrices $\Pi$
and $\Lambda$ (\ref{gor-pi},\ref{gor-l}) on the 3-dimensional case. It can allow as to obtain a
new three-dimensional natural  integrable system with higher order integrals of motion constructively.


\begin{thebibliography}{10}



\bibitem{en07}
Eilbeck J. C. ,  Enolski  V. Z.,   Matsutani S.,   \^{O}nishi Y., and  Previato E.:
\newblock{\em Abelian functions for trigonal curves of genus three},
 International Mathematics Research Notices, rnm 140-68, arXiv: math.AG/0610019, (2007).

 \bibitem{imm00}
Ibort A. ,  Magri F.,  Marmo G.:
\newblock {\em Bihamiltonian structures and St\"{a}ckel separability},
 {J. Geometry and Physics}, 33, 210-228, (2000).


\bibitem{ts99} Tsiganov A.V.: \newblock{\em Duality between integrable St\"ackel systems},
Journal of Physics A, Math.Gen., 32, 7965--7982, (1999).

\bibitem{ts00} Tsiganov A.V.: \newblock{\em	The Maupertuis Principle and Canonical Transformations of the Extended Phase Space}, Journal of Nonlinear Mathematical Physics, 8, 157-182, (2001).

\bibitem{ts10} Tsiganov A.~V.: \newblock{\em On bi-integrable natural
    Hamiltonian systems on the Riemannian manifolds}, arXiv:1006.3914, accepted to Journal of Nonlinear Mathematical Physics,   2011.

\bibitem{ts11s} Tsiganov A.~V.: \newblock{\em On natural Poisson bivectors on the sphere}, 	
J. Phys. A: Math. Theor., 44, 105203 (15pp),  (2011).

\bibitem{val10}
Valent G.: \newblock{\em On a class of integrable systems with a cubic first integral},
Commun. Math. Phys., 299, 631-649, (2010).

\bibitem{van05}
Vanhaecke P. : \newblock{\rm
Integrable systems and moduli spaces of rank two vector bundles on a non-hyperelliptic genus 3 curve},
Annales de l'institut Fourier, 55 (6),  1789-1802, (2005).


\bibitem{ts09v}
Vershilov A.V. ,  Tsiganov A.V.:
\newblock{\em
On bi-Hamiltonian geometry of some integrable systems on the sphere with cubic integral of motion}, J. Phys. A: Math. Theor. 42, 105203 (12pp), (2009).

\bibitem{w94}  Weierstrass  K.:\newblock{\em Mathematische Werke I}, vol. 1, 1894.

\end{thebibliography}
\end{document}